\documentclass{article}

\usepackage[english]{babel}

\usepackage[letterpaper,top=2cm,bottom=2cm,left=3cm,right=3cm,marginparwidth=1.75cm]{geometry}

\usepackage{amsmath}
\usepackage{graphicx}
\usepackage[colorlinks=true, allcolors=blue]{hyperref}

\usepackage{caption}
\usepackage{subcaption}

\usepackage[autostyle=true]{csquotes}

\usepackage[
backend=biber,
style=numeric,
sorting=none
]{biblatex} 
\title{Nanowire networks' interconnection graphs from their photomicrographs}
\author{J. Tau Anzoátegui, J.I. Diaz Schneider, E. Martínez, P. Levy, O. Filevich, C.P. Quinteros}

\begin{document}
\maketitle

\begin{abstract}
\noindent Self-assemblies of tunable units are being intensively studied as physical systems with signal processing abilities. Specifically, silver nanowire networks (AgNWNs) have demonstrated accumulation, non-linearity, and memory retention with multiple timescales, features that enable a wide variety of neuromorphic implementations. Abundant literature of either experimental or simulation studies analyzes the collective response of the systems, but attempts for linking the two strategies are scarce. In this study, we aim to extract the interconnection scheme to analyze the experimentally obtained network architecture and, eventually, be able to use it as the input of a previously developed simulation platform. By post-processing photomicrographs of AgNWN, we present a pipeline optimized to extract the interconnection diagram, recognizing the intersections formed among the nanowires, to determine the associated graph for each physical sample. A graph is a collection of nodes and edges whose properties can be associated with different electrical responses. It is thus possible to study graphs' metrics, such as the degree distribution, community size, \textit{clustering coefficient}, and \textit{path-length}, 
to compare the experimental assemblies' attributes to those of topological models of reference. \textbf{Small-world}, \textbf{modular}, and \textbf{scale-free} are well-known structures in the field of mathematical graphs. By analyzing the degree distribution, the adjacency matrices, among other useful representation means, the experimental assemblies reveal similarities to both \textbf{small-world} and \textbf{modular} topologies. The moderated degree values resemble the \textbf{small-world}-like networks, while the obtained community hierarchy and the sparser links between communities are more similar to \textbf{modular} graphs. All the mentioned analyses were conducted considering the interconnection scheme obtained from zenithal-view optical images, which overestimates the number of NWs interconnections (due to the impossibility of distinguishing real junctions from spurious cross-points between vertically displaced NWs). For that reason, this communication also studies the impact of artificially removing junctions from the resulting graphs on the previously calculated \textit{clustering coefficient} and \textit{path-length}. 

\end{abstract}

\section*{Introduction}

The computational strategies implemented by modern computers stem from concepts derived from the neurophysiology of biological brains \cite{haykin_neural_1999}. The encoding of information as memory states or trapped charge in synthetic units is one example (in hardware) of this \cite{jaeger_toward_2023}. The implementation of artificial neural networks composed of neuronal nodes and synaptic weights constitutes another example (this time in software) of the same original inspiration \cite{lecun_deep_2015}. However, the course of technological progress has distanced synthetic implementations from their biological counterparts \cite{palma-espinosa_balance_2025}. The power consumption of current computing systems and the restrictions this may entail in the near future necessitate proposing an alternative: exploiting synthetic systems radically different from the CMOS gates that are presently at the heart of the hardware \cite{jaeger_toward_2023}.

Self-assemblies of tunable units are being intensively studied as physical systems with signal-processing capabilities \cite{avizienis_neuromorphic_2012,milano_brain-inspired_2020,kuncic_neuromorphic_2021,rieck_ferroelastic_2023,jaeger_toward_2023,diaz_schneider_two-junction_2024,quinteros_thermal_2024}. Understood as complex networks of nanometric objects, these platforms (made of nanoparticles or nanowires, among others) resemble morphological, topological, and functional aspects of neural tissues \cite{palma-espinosa_balance_2025}. In that regard, the extremely high number of units, the spontaneous interconnection scheme among them, and the non-trivial architecture \cite{whitesides_self-assembly_2002} are desirable features for a new generation of neuromorphic hardware inspired by the collective attributes instead of the detailed fidelity of the individual constituents \cite{ariga_nanoarchitectonics_2021}.

In this framework, self-assemblies are being studied either experimentally \cite{avizienis_neuromorphic_2012,martin-garcia_solution-processed_2018,milano_brain-inspired_2020,kuncic_neuromorphic_2021,rieck_ferroelastic_2023,jaeger_toward_2023,diaz_schneider_two-junction_2024,quinteros_thermal_2024} or with simulation schemes \cite{bellew_resistance_2015,daniels_reservoir_2022,diaz_schneider_two-junction_2024}. Experimentally, synthesizing and measuring two or multi-electrode devices and subjecting them to a variety of incoming signals allows for the identification of desirable properties such as accumulation, non-linearity, short- and long-term memory, among others. Simulations performed on different platforms allow analysis of the impact of the properties of individual components (which are experimentally inaccessible) on the macroscopic collective response \cite{milano_connectome_2022}. Except for a few cases \cite{bellew_resistance_2015,rocha_ultimate_2015-3}, the interconnection schemes fed into the simulation platforms are not usually consistent with the experimentally obtained geometry. As such, the obtained simulated responses are just qualitatively comparable to the physical assemblies. In this study, we aim to bridge the gap between the two approaches by interpreting photomicrographs of the real samples as graphs of interconnected nodes useful to configure the simulations. Additionally, the conversion of the experimental information (optical images) into mathematical objects makes the latter suitable for network analysis in terms of their topology and geometry-related metrics.  

\section*{Synthesis and image acquisition}

\begin{figure}[ht!]
    \centering
    \includegraphics[width=0.5\linewidth]{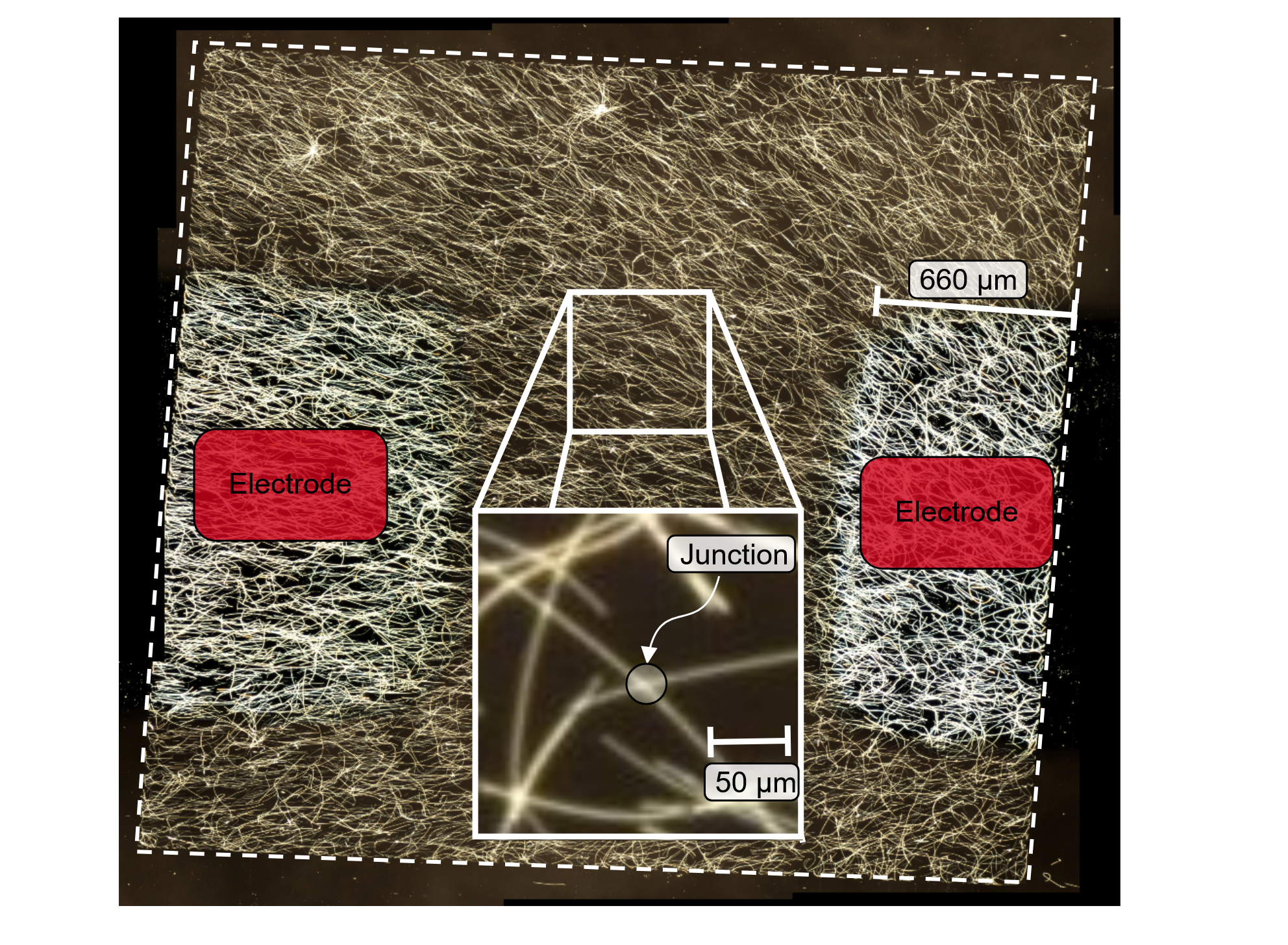}
    \caption{\textbf{Infograph of a typical nanowire network (NWN) depicting the constituent units.} Each NWN comprises the network of nanowires (NWs) intersecting each other, enclosed by two macroscopic electrodes placed 1 mm apart. A zoomed-in area is included as an inset where individual NWs and their junctions can be spotted. The image represents the so-called active area, which does not cover the full substrate surface but is a laterally delimited zone containing the NWN and the electrodes.}
    \label{fig:sketch}
\end{figure}

The AgNWs assemblies were fabricated with a targeted geometry ($\sim$ 170 nm in diameter and $\sim$ 70 $\mu$m in length), then dispersed in a solvent (containing polymeric residues coming from the NWs growth), and finally deposited onto a desired substrate to form the networks (AgNWNs) \cite{diaz_schneider_resistive_2022}. The desired areal density ($\sim$ 2500 mm$^{-2}$) is reached by sequential deposition steps \cite{diaz_schneider_resistive_2022}. As depicted in Fig. \ref{fig:sketch}, two millimetric-sized Ag electrodes (1 mm apart) were sputtered to access the assembly electrically.


\noindent The resultant assemblies comprise metallic NWs, coated with a polymeric layer residual from the synthesis, which intersect with each other, forming capacitor-like junctions (Ag-polymer-Ag). As such, the electrical connection between the two macroscopic electrodes is mediated by the conduction along the NWs themselves and across the tunable memristive junctions (indicated with an arrow in the inset of Fig. \ref{fig:sketch}). The electrical conductivity requires geometrical percolation of the assembly, but is also conditioned by the state of the memristive intersections among the constituent NWs. Fig. \ref{fig:sketch} shows the complete sample, including the nanowire network, NWN, deposited onto the solid substrate and the electrodes. Each assembly is a laterally delimited zone within its substrate referred to as the active area.    

Dark-field optical images were taken with a Leica DM2700 M using a typical magnification of 10x. Samples of intermediate areal densities $\frac{\mathrm{N}}{mm^2}$ (as reported in \cite{diaz_schneider_two-junction_2024}) were characterized using this technique. While in-plane scanning the sample, multiple images are taken to completely cover the assembly. These partial images are primarily stitched to obtain an image of the full active area. The image produced from this collection will be the input to the pipeline designed for the following graph extraction.    

\section*{Pipeline for graph generation}

\begin{figure}[ht!]
    \centering
    \includegraphics[width=\linewidth]{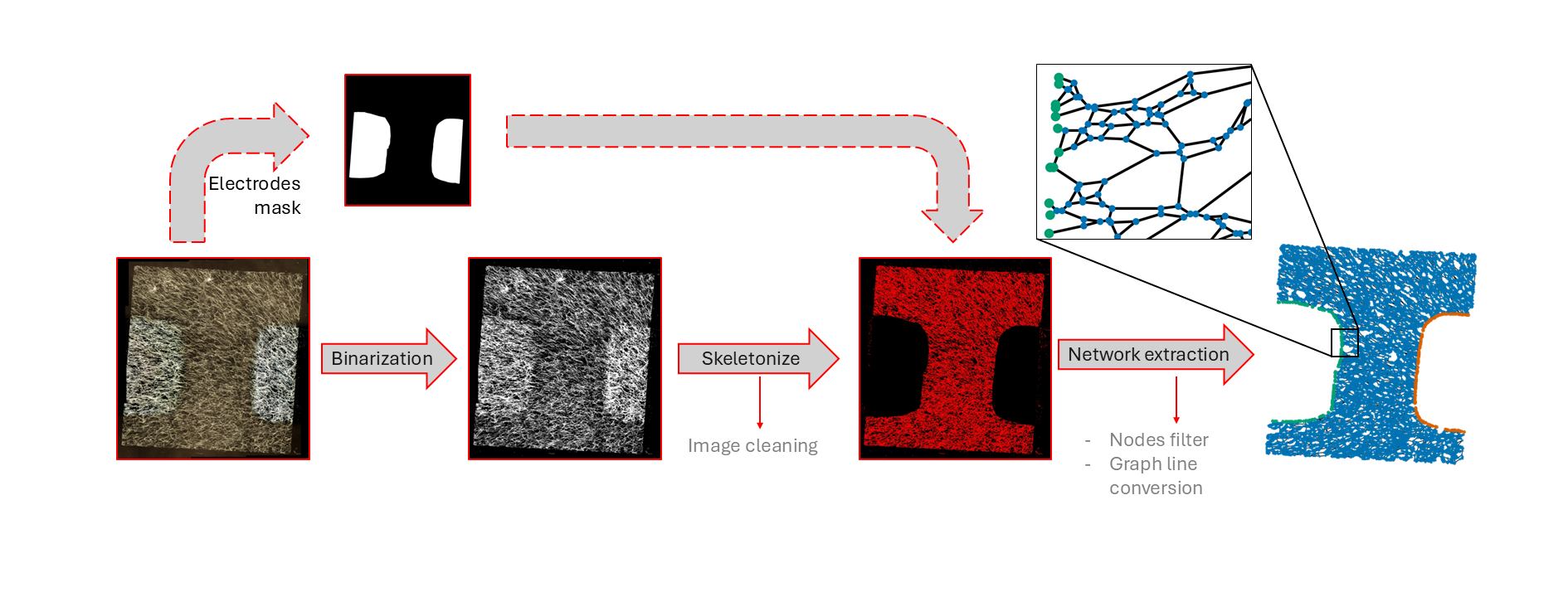}
    \caption{\textbf{Sketch of the pipeline flow.} Starting from the stitched picture, the procedure identifies the electrodes and binarizes the RGB-format image to facilitate the skeletonization conversion. Once the network has been converted to a set of one-pixel-wide contours, network extraction is performed. The wires and their mutual intersections (zenithal view) are interpreted as a graph comprised of nodes and edges, respectively, depicted in the zoomed-in area.}
    \label{fig:pipeline}
\end{figure}

The pipeline comprises a sequence of steps (schematically summarized in Fig. \ref{fig:pipeline}). First, the routine enables the identification of the macroscopic electrodes. This operation is of paramount importance since the graph objects contained within it will present a preferential hierarchy in the electrical percolation. This operation is customized, allowing for the user's flexibility in multiple electrode shapes and sizes. In the following, the NWN itself needs to be interpreted.    


\noindent Recognizing the nanowires and the intersections among them requires image processing strategies. In principle, binarization of the image facilitates posterior identification of the constituents. Additionally, an image cleaning procedure is implemented, where spurious pixels are filtered, removing both small black holes and isolated white noise. Once cleaned, the image undergoes a \textit{skeletonization} \cite{zhang1984} process to reduce the NWs to a single-pixel-wide line. This tool allows converting structures into a sketch of their contours. As such, the NWNs become a drawing of one-pixel-wide lines. The image obtained from the previous step allows for the intersection identification as the cross-points between lines. 

It is important to emphasize that a cross-point does not imply an intersection. From a zenithal view, it is not possible to determine whether two NWs occupying the same area and forming an angle between them are actually in contact (forming a real intersection) or if there is some vertical space between them. Parallax measurements would be required to disentangle the two cases. Unfortunately, the spacing between adjacent NWs may be so small that visible light's wavelength is not sufficiently short to provide such information. After building the corresponding graphs, this will be accounted for by introducing a removal probability. 

The result is a collection of segments and intersections. Since the segments are just portions of the original NWs (conductive and not tunable), while the intersections are the electrically active components, the segments will be coded as nodes and the junctions as edges. The reason behind this choice lies in the attribute assignment: edges are responsible for the weighted graphs, while the nodes represent the distribution of spots. Together, the number of nodes, 'N' (formerly segments of NWs), and of edges, 'e' (the intersections among NWs from the zenithal point of view) and their interconnectivity map will determine the type of structure associated with each experimental assembly. The number of links each node possesses is referred to as its degree (k). This definition can be extended to the network and represented either as an average calculated over all the nodes or as a distribution of values. 

Three samples (A, B, and C) were optically imaged and translated into graphs using the developed pipeline. The characteristics of the obtained graphs are summarized in Table \ref{tab:samples-graph-quantities}. Samples A and B possess similar densities, while C is sparser. The N-value is considerably higher for A and B than for C. However, N is not comparable to the number of NWs since the detection method breaks each wire into segments as determined by the cross-points with other neighboring NWs. N is not proportional to the total number of NWs either, since the number of segments depends not only on the total number of wires but also on the likelihood of intersecting other NWs, which increases with increasing NW density. The e-value is clearly related to the number of intersections with a monotonous dependence on the former. Interestingly, the average degree (net value) is similar for A and B and differs considerably for C. Nevertheless, more information needs to be collected in order to establish a relation between the density and the resulting nodes' degree.    

\begin{table}[ht!]
    \centering
    \begin{tabular}{|c|c|c|c|}
    \hline
    \textbf{Attribute} $\downarrow$ / \textbf{Sample} $\rightarrow$ & A & B & C \\
    \hline
    \hline
    Nodes (N) & 66435 & 77477 & 26222 \\
    \hline
    Edges (e) & 121217 & 139293 & 41941 \\
    \hline
    Degree (average) & 3.65 & 3.60 & 3.20 \\
    \hline
    \end{tabular}
    \caption{Graph features of the experimental samples.}
    \label{tab:samples-graph-quantities}
\end{table}

\section*{Topological analysis}

Graph generation from experimentally-acquired optical images enables geometrical considerations. Multiple metrics are defined as indicators of the network segregation and/or integration and are related to some archetypal interconnection schemes. Among them, we could mention: \textbf{small-world}, \textbf{modular}, and \textbf{scale-free} networks. They differ in their main characteristic, which affects their architecture. \textbf{Small-world} networks \cite{watts_collective_1998} are typically formed by clusters, within which the connectivity is high, and hubs linking nodes belonging to different clusters. In this way, two metrics are defined: \textit{clustering coefficient} (CC) and \textit{path length} (PL). 

The \textit{clustering coefficient} defines the likelihood that a graph is formed by subgroups. Specifically, the local \textit{clustering coefficient} of a node $u$ is the fraction of possible triangles through that node that exist \cite{networkx_clustering, onnela2005}:
\begin{equation}
    \mathrm{CC} =\frac 1N \sum_u c_u=\frac 1N \sum_u \frac{T(u)}{\deg(u)(\deg(u)-1)}
\end{equation}
where $T(u)$ is the number of triangles through node $u$ and $\deg(u)$ is the degree of $u$. Since the macroscopic topology of the system is the primary focus, we evaluate the average \textit{clustering coefficient}, defined as the mean of $c_u$ over all $N$ nodes in the graph. Hereafter, any mention of the \textit{clustering coefficient} will refer to this global average CC.

The \textit{path length} is defined as the mean of the minimum distance (in number of edges) between all pairs of nodes in the network: 
\begin{equation}
    \mathrm{PL} =\frac 1{N(N-1)} \sum_{u\neq v}d(u,v)
\end{equation}
where $d(u,v)$ is the shortest distance between node $u$ and $v$. While this distance can be calculated through various algorithms, computing the exact distance between every possible pair incurs a prohibitive computational cost for large-scale graphs. To overcome this, a statistical approximation was implemented. First, the analysis was restricted to the largest connected component of the graph, as the distance between disconnected nodes diverges to infinity. Within this main component, a representative subset of $m = 1000$ random seed nodes was sampled. Using sparse matrix algorithms to compute the shortest paths, the minimum distances from these seed nodes to all other nodes in the component were evaluated. The final PL value reported corresponds to the arithmetic mean of this subset of shortest paths, providing a robust estimator of the global macroscopic distance.

\textbf{Small-world} networks, early proposed by Watts-Strogatz \cite{watts_collective_1998}, possess a short average \textit{path-length} and a high \textit{clustering coefficient}. In turn, \textbf{modular} networks \cite{girvan_community_2002} are characterized by presenting communities that determine a coexisting structure of sparse and dense connectivity zones. Adjacency matrices are usually used to illustrate this behavior. They are two-dimensional graphical representations that order the nodes in vicinities and color the intersections between linked nodes. The diagonal of such a matrix represents identity (each node is by default linked to itself), while the squares formed around represent the communities formed by profusely linked nodes. The specificity of the modular networks is that the size of those communities is very well defined.  
Finally, \textbf{scale-free} networks \cite{barabasi_emergence_1999} have a highly heterogeneous connectivity distribution, with most nodes having very few links and a few nodes (identified as hubs) linking to an exceptionally high number of nodes. By construction, these networks possess multiple spatial length scales.

\begin{figure}[ht!]
    \centering
    \includegraphics[width=.9\linewidth]{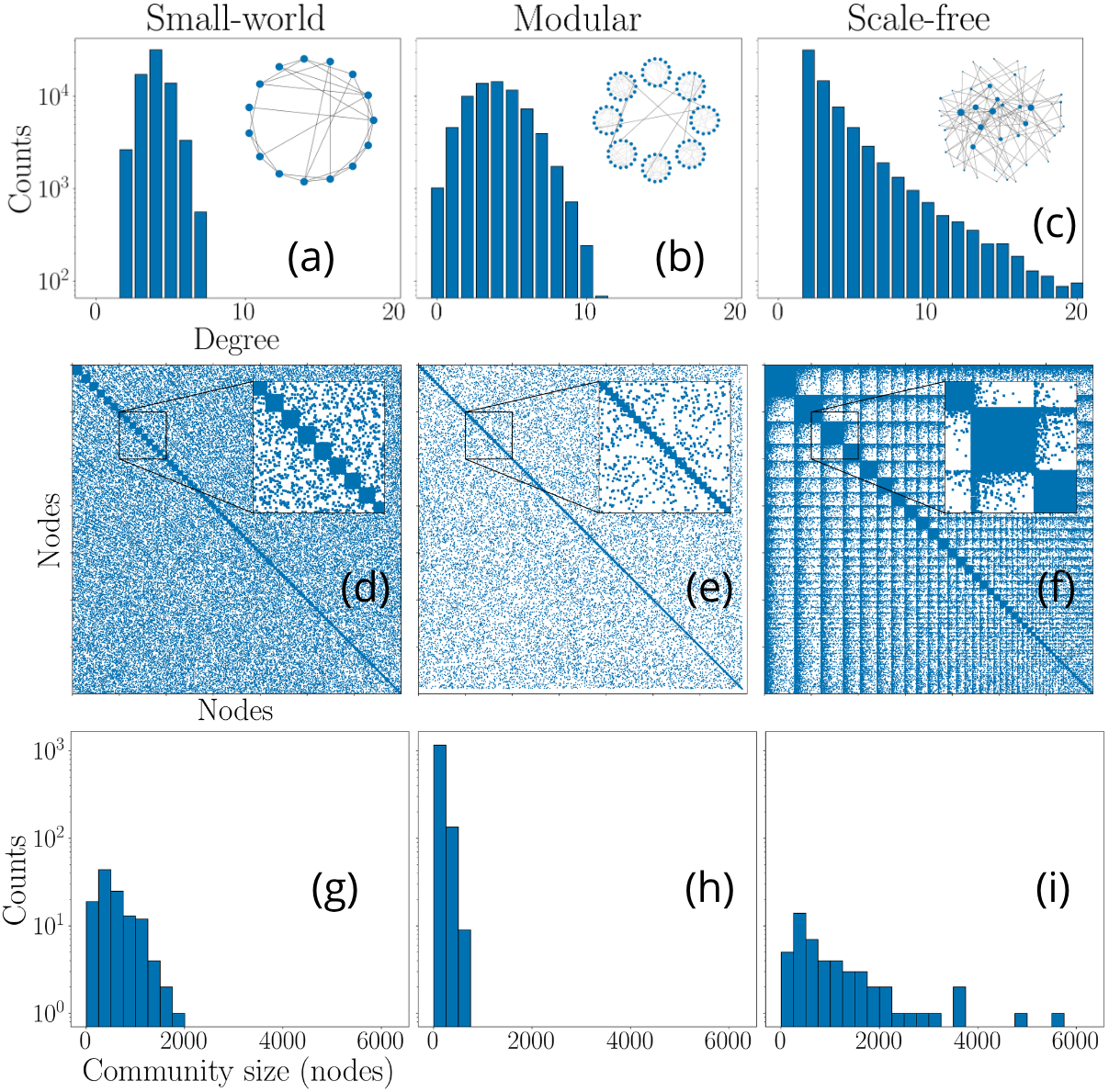}
    \caption{\textbf{Different network schemes used as a reference: small-world, modular, and scale-free.} The top row, (a), (b), and (c), comprises a degree histogram including an inset with sketches of a moderate size to help visualization. On the second row, (d), (e), and (f) show the adjacency matrix, while the bottom row, (g), (h), and (i), displays the distribution of community sizes for each case. Degree distributions, adjacency matrices, and their associated community size distributions are calculated for networks of a size and average degree comparable to the experimental realizations (N = 70000, k = 4).}
    \label{fig:reference-networks}
\end{figure}

\noindent Although the three archetypal networks are not mutually exclusive, in the following, we will use them as topological references to compare the graphs obtained from the optical images using the developed pipeline. Fig. \ref{fig:reference-networks} represents some of their salient features, quantified in artificially built networks generated with size N = 70000 and average degree k = 4, which in turn is comparable to the amount obtained for the experimental cases. Figs. \ref{fig:reference-networks}(a), (b), and (c) represent the degree distribution for a \textbf{small-world}, a \textbf{modular}, and a \textbf{scale-free} network, respectively. For each case, a sketch is included as an inset for visual reference. Figs. \ref{fig:reference-networks}(d), (e), and (f) comprise their corresponding adjacency matrices, allowing us to identify the presence of communities as the dark squares surrounding the matrix diagonal and sparser connections among those communities depicted as colored regions outside the diagonal. The community sizes may vary (see the insets in (d), (e), and (f)), and their distributions are shown in Figs. \ref{fig:reference-networks}(g), (h), and (i). The latter help to identify at a glance the community size distribution in each case. While the \textbf{modular} network depicts a well defined value, the other two possess wider distributions, indicating the existence of communities with a broader size range.

\begin{figure}[ht!]
    \centering
    \includegraphics[width=0.9\linewidth]{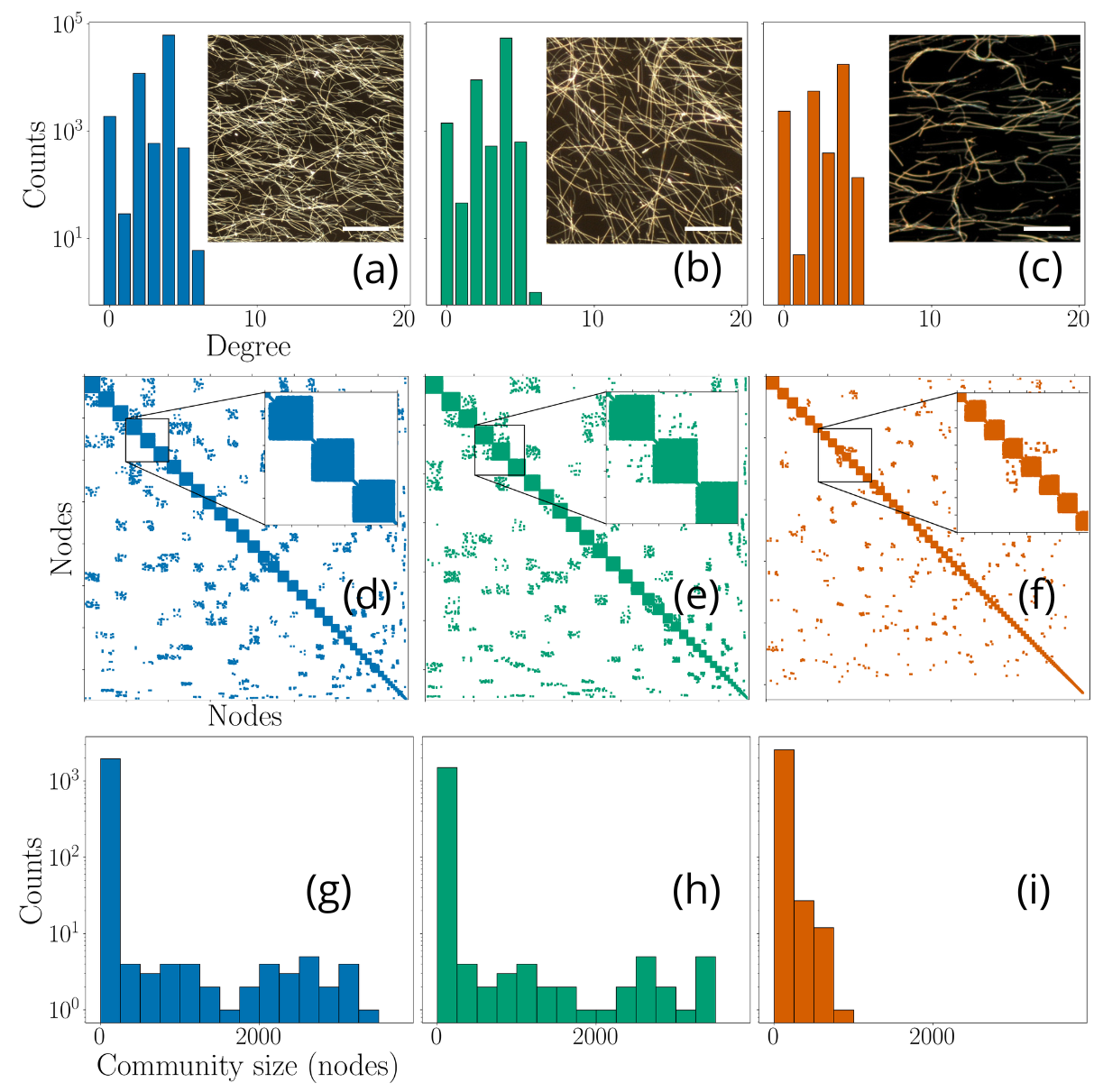}
    \caption{\textbf{Network metrics of graphs generated from experimental images.} Degree distributions with optical images as insets with a scale bar of 100 $\mu$m ((a), (b), and (c)), adjacency matrices ((d), (e), and (f)), and community distributions ((g), (h), and (i)) for three experimental cases (samples A, B, and C, respectively).}
    \label{fig:experimental-histograms-matrices}
\end{figure} 

\noindent Fig. \ref{fig:experimental-histograms-matrices} depicts the degree distribution and adjacency matrix obtained from three different samples: A, B, and C. Those labeled as A and B have the same nominal areal density, while C is smaller. This difference is noticeable in Figs. \ref{fig:experimental-histograms-matrices}(a), (b), and (c) insets. Comparison of the obtained graph characteristics corresponding to the experimental samples allows for identifying a moderate range of available connectivity degrees, which is compatible with \textbf{small-world}-like schemes. However, adjacency matrices show a \textbf{modular} arrangement as indicated by the presence and sustainable size of the squares forming around the matrices' diagonals. Additionally, the spots appreciated outside the diagonal (symmetric relative to the diagonal) indicate extra-community interconnections.


As mentioned in the Introduction, the experimentally available zenithal view does not enable distinguishing real junctions from concurrent NWs that may not be in physical contact (they are imaged within the same window but vertically displaced, i.e. not in physical contact). So far, all the intersections have been considered as effective junctions and reflected in the graph as memristive edges. This is not realistic. However, since it is not straightforward to experimentally implement a strategy to differentiate the two cases, here we explore the graph variability by quantifying some of the previous metrics as a function of an arbitrary edge-removal probability (p$_{\mathrm{er}}$). To incorporate this, the graph is modified by scanning every node and applying a previously defined p$_\mathrm{er}$. Given the statistical significance of the number of constituents, the obtained graph is expected to have an amount of survival edges proportional to (1 - p$_{\mathrm{er}}$).


\begin{figure}[ht!]
    \centering
    \includegraphics[width=\linewidth]{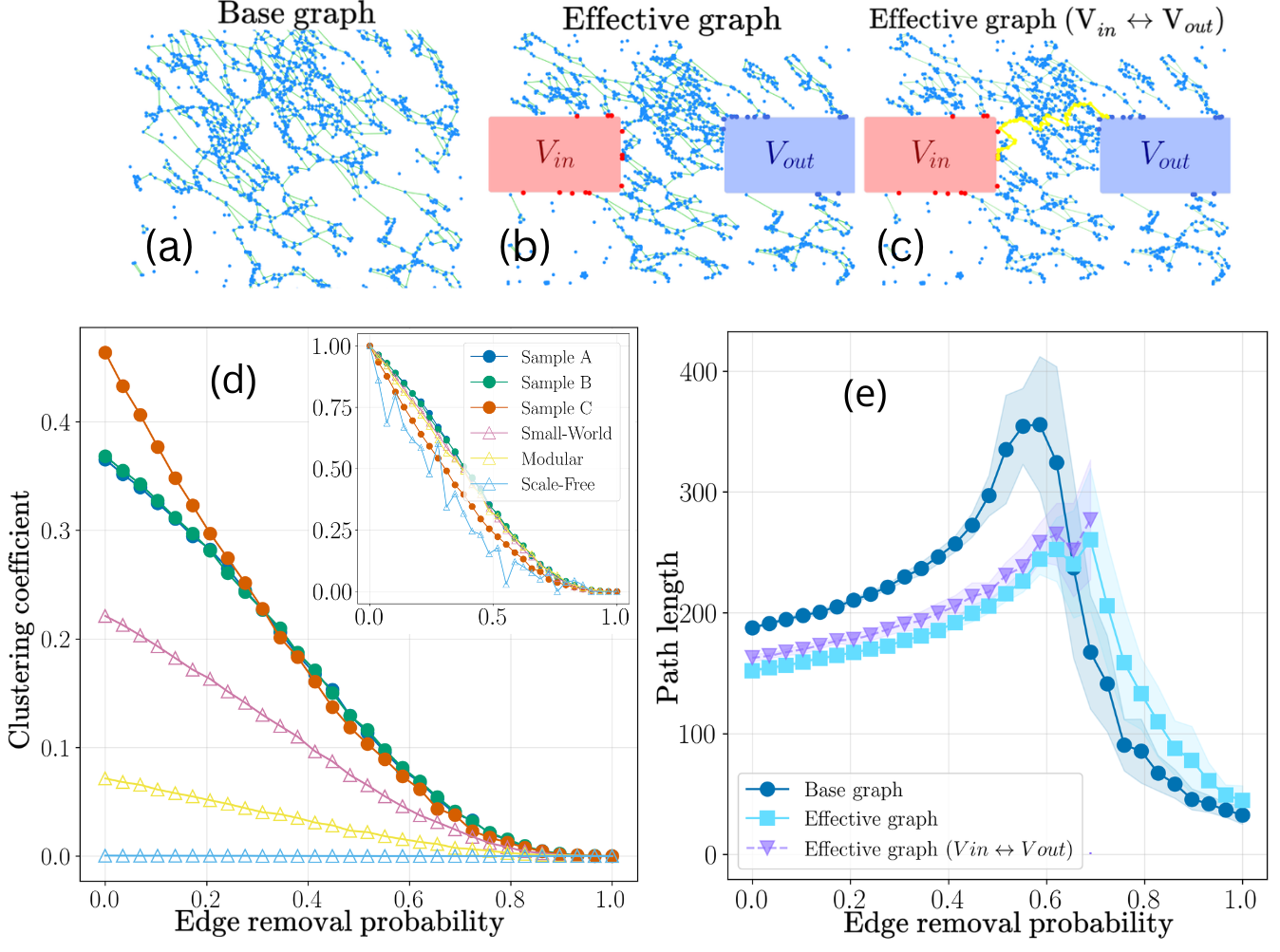}
    \caption{\textbf{\textit{Clustering coefficient} and \textit{path length} as a function of the edge-removal probability.} (a), (b) and (c) depict schematics of the different graphs used to measure the \textit{path length}. (d) shows the \textit{clustering coefficient} as a function of the edge-removal probability for the samples and the reference networks with an inset showing the \textit{clustering coefficient} normalized by the initial value of each graph. (e) displays the \textit{path length} as a function of the edge-removal probability for Sample A in the three different cases shown above.} 
    \label{fig:removal-probability}
\end{figure}

\noindent Zenithal view of the optical images prevents distinguishing between real junctions (tunable intersections between two NWs) and just crossing NWs that are concurrent but vertically shifted to each other (false junctions). Considering p$_{\mathrm{er}}$ is a strategy to afterwards modify the obtained graph, such that the impact of false junctions on the graph metrics can be tackled. \textit{Clustering coefficient} and \textit{path length} are displayed as a function of p$_{\mathrm{er}}$ (Fig. \ref{fig:removal-probability}). 

Starting from the \textit{clustering coefficient} determined for the whole network as if all the cross-points were real junctions, removing edges produces a reduction of the agglomeration capability. The more edges are statistically erased, the smaller the \textit{clustering coefficient} becomes at a close-to-fixed rate until the amount of remaining edges is so small that the approach to zero becomes asymptotic. There is no dramatic difference among the three analyzed samples, although it has to be considered that the three of them were very similar to each other in their initial condition. In this sense, the same analysis was conducted with the three networks of reference, showing that despite the initial condition being different for each case (see Fig. \ref{fig:removal-probability}(d)), the overall evolution as a function of the edge-removal probability is the same (inset of Fig. \ref{fig:removal-probability}(d)). 

The \textit{path length}, Fig. \ref{fig:removal-probability}(e), deserves a more detailed explanation. Since it relates more closely to the expected electrical response of the network (not analyzed in this study), this quantity has been calculated in three different ways. To account for the inherent variability of this process, the data for each of these three methods were averaged over 20 independent runs, with the shaded regions corresponding to the standard deviation. The first one considers the \textbf{\textit{base graph}}, which is the interconnection scheme identified regardless of the electrodes (see Fig. \ref{fig:removal-probability} (a)). The second one considers that the nodes located within the electrodes' areas are equivalent to each other since they short-circuit an equipotential surface; this is named after \textbf{\textit{effective graph}}. The last one (\textbf{\textit{effective electrode graph}}) redefines the \textit{path length} definition considering that the only paths to be considered are those of electrical validity, meaning that the only possible origin nodes are those belonging to the V$_{\mathrm{in}}$ electrode and the only possible destination nodes are the corresponding to V$_{\mathrm{out}}$. This limits the combinatorics of the eligible paths to be considered. Specifically, this constraint forces the discarding of some of the largest paths, considering that the furthest nodes (i.e., those lying along the diagonal of the active area) are not summed up. This is the reason why the net values of the \textit{path length} are shifted down in the \textbf{\textit{effective graph}} compared to the \textbf{\textit{base graph}}. Moreover, upon edge removal, the \textit{path length} increases due to the absence of nodes' links that previously could have worked as intermediaries. Eventually, the edge removal is so high that part of the graph becomes disconnected (meaning that not every node pair is physically connected any longer). The maximum removal probability can be thought of as the inverse of the graph's minimum density to be fully connected, which is equivalent to the percolation limit. Calculating the \textit{path length} in the \textbf{\textit{effective graph}} is still possible by discarding those pairs that are not connected. The \textit{path length} calculated within such a disconnected graph is performed among those pairs still connected, with a higher likelihood for closer nodes than for further ones. Finally, in the \textbf{\textit{effective electrode graph}}, the \textit{path length} is not measurable beyond the percolation limit since there is no remaining path satisfying the imposed restriction.    
The topological analysis of the pipeline-generated graphs indicates that the experimental assemblies do not strictly correspond to just one type of archetypal network architecture, but hints of the three types could be found. In addition, the metrics calculated as if all the cross-points were real junctions are defective, and this is considered here by introducing an arbitrary removal probability. This strategy reveals that the obtained values might differ upon introducing a more realistic tridimensional connectivity map. Despite this, which requires further experimental improvement, the conducted analysis reveals the usefulness of the developed pipeline and the validity of the geometrical analysis it enables.     

\section*{Conclusions}

This work presents a pipeline developed to build a graph or network as a mathematical object associated with the optical images of experimental samples. 
Graphs' attributes, such as the number of obtained nodes and edges, reflecting NWs' segments and intersections, respectively, were compared among three experimental samples. Moreover, topological metrics and means of representation derived from three networks chosen as references were used to relate the obtained graphs to well-known structures. Considering that all the identified intersections reflect real junctions, all the resulting graphs displayed a degree distribution similar to a \textbf{small-world} distribution. In turn, the adjacency matrices showed the presence of communities of different sizes, which are not mutually isolated but interconnected. 
Other topological metrics, such as the \textit{path length} and \textit{clustering coefficient}, were calculated not only for the \textbf{base graph}, considering the whole active area and all the zenithal intersections as real junctions, but also as a function of an artificially introduced edge-removal probability. The latter accounts for the possibility of the intersections not representing physical contacts between the NWs and, consequently, constituting false junctions. 
This artificial insert, aimed at accounting for the realistic status of the NWN, allowed us to test the impact of the effective density on the metrics. More importantly, it reinforces the idea that not only does the graph, understood as the nodes and links distribution, impact their calculation, but also the shape and area of the electrodes used to electrically access the network. Future work will be devoted to further exploring the impact of the electrodes' definition on the topology, but also to testing multiple electrical protocols using previously developed simulation platforms.       

\section*{Acknowledgments}

The authors kindly acknowledge Prof. Zdenka Kuncic, Ms. Trinidad Ibar, Dr. Victoria Rosato Siri, Dr. Federico Golmar, Dr. Lucas Finazzi, and the members of the LINE group and IA-CoNSoFi consortium for their insightful comments and the fruitful discussions. This work was partly supported by CONICET PIP 2023-2025 11220220100508CO and CONICET PIET-R 2025 29820250100057CO.

\printbibliography

\end{document}